# Cell Development obeys Maximum Fisher Information

B. Roy Frieden[1], Robert A. Gatenby[2]

[1]College of Optical Sciences, University of Arizona, Tucson, Arizona 85721

[2]Departments of Radiology and Integrated Mathematical Oncology, Moffitt Cancer Center, Tampa, Florida 33612

**TABLE OF CONTENTS**

*Keywords*

# 1. ABSTRACT

Eukaryotic cell development has been optimized by natural selection to obey maximal intracellular flux of messenger proteins. This, in turn, implies maximum Fisher information on angular position about a target nuclear pore complex (NPR). The cell is simply modeled as spherical, with cell membrane (CM) diameter *10*μm and concentric nuclear membrane (NM) diameter *6*μm. The NM contains ≈ 3000 nuclear pore complexes (NPCs). Development requires messenger ligands to travel from the CM-NPC-DNA target binding sites. Ligands acquire negative charge by phosphorylation, passing through the cytoplasm over Newtonian trajectories toward positively charged NPCs (utilizing positive nuclear localization sequences). The CM-NPC channel obeys maximized mean protein flux $F$ and Fisher information $I$ at the NPC, with first-order $\delta I = 0$ and approximate 2nd-order $\delta^2 I \approx 0$ stability to environmental perturbations. Many of its predictions are confirmed, including the dominance of protein pathways of from 1-4 proteins, a *4nm* size for the EGFR protein and the flux value $F \approx 10^{16}$ proteins/m$^2$-s. After entering the nucleus, each protein ultimately delivers its ligand information to a DNA target site with maximum probability, i.e. maximum Kullback-Liebler entropy $H_{KL}$. In a smoothness limit $H_{KL} \to I_{DNA}/2$, so that the *total* CM-NPC-DNA channel obeys maximum Fisher $I$. Thus maximum information → non-equilibrium, one condition for life.

# 2. INTRODUCTION

## 2.1 Role of ligands; prescriptive information

An essential step in cell development is when an extracellular messenger ligand arrives at a lateral angular position $x_0$ on the cell membrane (CM) of a given cell. The ligand binds to a protein there [1]. For simplicity, the CM is modeled as a sphere of diameter *10*μm, containing a concentric nuclear membrane (NM) of diameter *6*μm with multiple nuclear pore complexes (NPCs). The medium between CM and NM consists of cytoplasm. The latter, consisting of cytosol (70%), organelles, protein membranes, ions, lipid droplets, etc., is modeled, for the sake of calculation, as an approximately homogeneous medium, with the same mass density ρ, drag coefficient and scattering variance $\sigma^2$ throughout. Some workers call the information that is carried by the ligand "prescriptive," in the sense of providing biological instructions for a target DNA codon site to follow. These are conducive to overall cell development or maintenance. Such prescribed information (PI) is manifest in a particular arrangement of molecules to be carried by the ligand, e.g. an enzyme. We do not define the message contents any further. See, e.g., [2] for further definition. Although it can be argued that there is no evidence for attributing any prescriptive information to such ligands, i.e. they have zero biofunction, there are arguments to the contrary as well [3].

## 2.2 A role of natural selection

We assume that, by natural selection, the DNA has selected for ligand paths that deliver the information efficiently, that is spatially accurately and with minimal time to its intended DNA site. For this purpose, the DNA has directed the creation of a strong enough electric field to efficiently direct each ligand to its target NPC site on the nuclear membrane as described below.

To accomplish this, the ligand is first scaffolded for phosphorylation. Its resulting negative charge promotes motion toward the target DNA site through, first, Coulomb attraction toward an intermediary, positively charged NPC on the NM.

## 2.3 The two information channels

Thus the trajectory of the ligand is from CM-NPC-DNA. This is analyzed in two steps: as a CM-NPC channel (Secs. 3-8), and then an NPC-DNA channel (Sec. 9) Note that the vast majority of the analysis is of the CM-NPC channel, i.e. exterior to the DNA content of the genome. Our overall objective is to predict how accurately the ligand information can be delivered to its final destination the DNA, and also to determine if the predictions are confirmed by laboratory observation.

## 2.4 Role of Fisher information

The role of Fisher information in all this is to force, with high probability, the messenger protein to enter its target NPC; and then, once it enters, to travel to and alter the DNA codon that the ligand message prescribes. Maximizing the Fisher information on angular position of the ligand will be seen to foster this effect. See Sec. 4.1 and Appendix B.

**2.5 Essential differences between two entropies S and H**

We shall often refer to the system 'entropy,' so at this point it is important to clarify the concept. There are two closely-related entropies, the Shannon (or 'information') entropy $H$ and the thermodynamic entropy $S$. The latter is the entropy referred to in the Second law of thermodynamics. The relation is that $S = kH$, where $k$ = Boltzmann's constant). Another comparison lies in their meanings: information $H$ measures epistemological uncertainty in the occurrence of system events; whereas thermodynamic $S$ is a physicodynamic measure, specifically of heat or some other system waste product (We thank a reviewer for emphasizing this distinction.)

**2.6 Basic differences between Fisher information and entropy H**

There is also an important distinction to be made between the concepts of *Fisher information and the Shannon entropy H*. Fisher information measures the level of structural order, or complexity, of a system; whereas both $H$ and $S$ (since they are proportional) measure the resulting amount of waste entropy given off as the price of forming the order. For example, the level of internal structure of a horse is measured by its Fisher information, rather than by the amount of waste entropy it excretes in the process of growing that structure. *You can't reconstruct the horse by minute examination of its manure*.

A further indication is that Fisher information has *physical units* (since it is information *about something real*) whereas Shannon does not. These factors were anticipated in [4] That author points to the warning given by Brillouin, himself, that Shannon entropy $H$ ignores the value or the meaning of the information which is quantified by the definition.

**3. INFORMATION IN CM-NPC CHANNEL**

Let the angular position of the input ligand on the CM be $x_0$. It is reasonable that 4 billion years of natural selection, normal cell development delivers the ligand to the NM in *minimal time*. Minimal time is crucial when a cell suffers a sudden trauma such as a wound or incursion by a foreign body such as a virus. Note that this argument *assumes* that the ligand has specific biofunction, as mentioned above. We do not attempt to derive its biofunctional aspect here. We implicitly assume that it has developed in parallel with the assumption here of efficient delivery of the information to the DNA site. As mentioned below, it is the 'death' aspect of evolution that tends to allow those cells that have such advantageous information properties that have higher fitness than, and therefore dominate in population over, those that do not.

This permits, e.g., quick response to sudden trauma such as wounds. To accomplish minimal time, *the ligand should travel radially*, i.e. toward the center of the cell. (See Sec. 7.1 for further details.) Thus *the ideal angular position on the NM is likewise $x_0$*. However, as with any real information channel, it suffers inevitable noise $x$ of random angular displacement. Here the noise is diffusion due to random collisions with particles of the cytoplasm. Hence the messenger protein arrives on the NM at a generally *non*ideal positional data value $y = x_0 + x$, with mean $<x>=0$. Denote the probability density function (pdf) on the noise from NPC position $x_0$ as $p(x)$.

**3.1 Fisher information**

The information $I$ about ideal NPC position $x_0$ that is carried over this CM-NPC channel by the messenger protein is

$$I = \int dx \frac{\left(\frac{dp}{dx}\right)^2}{p}, \quad p = p(x), \text{ with } e_{min} = 1/\sqrt{I}. \quad (1)$$

Quantity $I$ is called the 'Fisher information' [5]. Fisher information is always information 'about something.' In this case that something is the ideal position $x_0$. Note also its significance in Eq. (1) of fixing $e_{min}$, the minimum possible mean-square error from ideal position $x_0$ at the over the many proteins (for example, ≳ 400,000 molecules per protein type in a typical liver cell) transiting the channel $x_0 \rightarrow y$. The 3rd Eq. (1) shows the reasonable result that the higher the information is the lower is the minimum mean-square error.

Aside from this significance as a measure of data quality, *I* also has the significance of defining the degree of order or *complexity*, Eq. (18), of a system (see also Appendix A) Thus Fisher information is a measure of both *data quality* and *system quality*. Such directed positioning could be considered another dimension of Prescriptive Information, in particular regarding the question of how the cell is able to position ligands for entropy into the target nuclear pore. Finally perturbations in *I* due to coarse graining events (see below Eq. (18)) define a biological *arrow of time* (Appendix A).

In fact, how transport through the pore is achieved might be considered another informational question relevant to protocell or primordial cell formation. The formative mechanism was death: Those cells that did not consistently achieve efficient or near-efficient transport would have been eliminated by natural selection.

### 3.2 Principle of maximum Fisher information

Since NM location errors *x* are departures from the *minimum possible (radial) distance* traveled by the protein from CM-NPC, by our working hypothesis of minimal CM-NPC travel time the error $e_{min}$ should be minimal [6a-8a]. That is, by Eq. (1), messenger proteins take intracellular paths that *maximize their levels of spatial Fisher information I* at the NM,

$$I = max. \qquad (2)$$

As we discuss in Sec. 5.1, due to Eq. (3) these also maximize the protein flux at the NM and also minimize the trajectory time, i.e. reaction time, to the DNA target as required after trauma.
Eq. (2) is our working hypothesis. It was originally used in deriving laws of inanimate systems [6a], and from the fact that for a particular living system – cancer – the Fisher is found to be *minimized* [6a,b]. Since cancer it randomness personified then any normally developing system (as assumed here) is assumed to be developing in the *opposite* direction – toward *maximum order* or Fisher information.

A final need for maximum positional information arises out of the needs of morphogenic signaling. In developmental biology, morphogenic gradients direct organ and tissue formation in fetal development. This requires normal cells to recognize and accurately measure a gradient of morphogens across its diameter; see Example section in [8b].

Information level (2) captured by a population also leads [7] to Fisher's fundamental theorem of natural selection.

Likewise, in protocell formation such accurate positioning would foster optimally morphogenic development, again with disease or death as the punishment for not accomplishing it well enough. As a consequence, the genetic code of such cells would dominate over the less able. Note that *I* is information about the environmental factors entering into the determination of the Debye-Huckel parameter $k_0$ in Eq. (4). That is, the protein density $\rho$, their charge $q$, the cytoplasm temperature $T$ and its dielectric constant $\varepsilon$.

We may note in passing that *I* = max. is likewise a property of inanimate matter, again, for finite-sized systems; note *finite* extension *L* in Eq. (18). That I = max. does not violate the 2$^{nd}$ law of thermodynamics; the latter requires order to decrease universally, but allows it to increase locally, i.e. within finite-sized systems. Thus, it is notable that principle (2) leads as well to derivation of the laws of *inanimate, physical* systems [8a]. Thus the application of (2) to biological systems follows on the grounds that they are, ultimately, *physical on the microlevel*. It is undeniable that living systems, such as the DNA backbone, consist of chiral molecules that are the mirror images of those currently occurring naturally; but there is no principle stating that such enantiomers cannot occur under proper physical conditions. Undoubtedly, someday such conditions will be found.

Nevertheless life has characteristics that uniquely differ from those of current non-life systems, e.g. the ability to reproduce by simple mitosis. However, again, this does not mean that eventually such systems could not be artificially produced.

Regarding why condition Principle (2) has to be imposed to allow life to persist, it is because, as explained below, random diffusion by the molecules of the cytoplasm would otherwise widely divert the messenger proteins from their target NPCs. That is, in the general randomness of the environment and its incursions into the living system have to be kept under control in order to achieve the target NPC and, then, target DNA site. This also turns out to keep the whole system stable to 2$^{nd}$ order perturbation (as found below).

As important checks, principle (2) makes predictions about cell biology that are confirmed by lab observation [8b]: dominance of two-three component protein pathways (RAS, RAF, MEK, ERK, etc.), stability of cells, prediction of very high E-field strengths confirmed by the Kopelman group [9], the central role played by phosphorylation, a very fast protein response time, the order of 0.016s, and an optimal messenger protein size of about 4 nm.

Regarding whether Fisher information is prescriptive or descriptive, it is both. It is prescriptive since, in being forced to obey principle (2), it forces cell entropy to be minimal (see below), therefore as far as possible from 'dead,' while simultaneously forcing the cell to remain in equilibrium despite the presence of perturbations up to 2$^{nd}$ order size. It also is descriptive in giving rise to the above (and other) requirements on the performance of a cell.

By the 3rd Eq. (1), attaining maximum information value (2) results in minimum root mean-square (rms) transverse positional error $e_{min}$ on the NM (thus, a minimum *of a minimum* value). Hence NM position $x_0$ can, as in an optical system, be regarded as the geometrical conjugate position to its launch position on the CM. Accomplishing such (doubly) minimal error also results in a *minimal travel time* $t_a$ from CM to NM (our aim, stated in Sec. 3). This allows a version of "time gating," to be accomplished by each NPC on the NM, with the aim of selectively allowing only *required* proteins to enter. "Time gating" is taken up in Sec. 7.3.

This paper both summarizes the work [6b, 8a, 8b] on cell development resulting from principle (2) and gives new results that follow from it. It is shown in Appendix D that this information is proportional to the mean flux $F$ (number/area/time) of proteins within the cell cytoplasm (CM-NPC) channel, obeying Eq. (D2),

$$I = \left(\frac{A}{2D}\right) F. \qquad (3)$$

Area $A = \pi a^2 = 28.3 \mu m^2$ (since cell radius $a = 5$ μm) is the cross sectional area of the CM, and diffusion constant $D = 5 \times 10^{-11}$ m$^2$/s of the cytoplasm. The flux $F$ values have been calculated [8a], as taken up below, so Eq. (3) allows the information $I$ to be quantified.

### 3.3 Debye-Huckel constant

A parameter of the cell that is key to determining its ability to transport proteins is the Debye-Huckel constant $k_0$ of its cytoplasm. This obeys [8a]

$$k_0 = \sqrt{\rho q^2 (\varepsilon k_B T)^{-1}}, \qquad (4)$$

where $\rho$ is the mass density of proteins in the cytoplasm, $q$ is the electric charge on each, and $\varepsilon, T$ are the dielectric constant and temperature, respectively, of the cytoplasm.

Typical such transport proteins are those in the RAS, RAF, MEK or ERK protein channels. The approach would generally give a different answer for the flux $F$, information $I$ and traversal time $t_a$ of each protein type. For example, the drag force on the protein depends upon its length. And the latter varies from one protein type to the other. Our aim is to find typical answers for a typical protein, as defined by the parameters in Table 1 below.

However, there is a tendency for generality of the results, in that the acceleration term $ma$, with $m$ = protein mass, $a$ = acceleration, contributes negligibly to the total force on the protein. (See line preceding

Eq. (B3) in Appendix B.) Therefore proteins with different masses still obey the answer we obtained since, from the above, only their lengths matter. The physical reason for the mass drop-out is that the acceleration term *ma*, *a* = acceleration, contributes negligibly to the total force on the protein. The biological reason is that it would be unreasonable to imagine such a large force of acceleration on the protein. So the solution is that it effectively travels at (instantaneous) "terminal velocity" from CM to NM, almost like a particle falling through space in free-fall. However, this velocity will depend upon the instantaneous drag force and, hence, the protein length. The latter affects the Reynolds number given in Table I immediately below.

Perturbations in the parameters ρ, q, ε and *T* defining $k_0$ in Eq. (4) will be shown [below Eq. (10)] to have a *remarkably small* effect on perturbing *F* and therefore, by (3) and (1), on perturbing *I* and error $e_{min}$. The size of $k_0$ defines the degree to which proteins in the cytoplasm are shielded from electric fields. The model assumes the particular cell parameters in Table 1, meant to approximate those of a typical human myocite cell. The number 462 defining the Reynolds number is the number of amino acids comprising a typical human messenger protein.

**Table 1.** Cell parameters used in the calculation

| | |
|---|---|
| CM radius $r_0$ | 5 μm |
| NM radius *a* | 3 μm (Note: $a/r_0 \approx$ 60% for mammalian cells) |
| Cytoplasm dielectric const. $\varepsilon = 60\varepsilon_0$ | $7.1 \times 10^{-10}$ F/m |
| Cytoplasm drag coefficient *D* | $0.27 \times 10^{-11}$ m$^2$/s |
| Thermal energy $k_B T$ | $4.14 \times 10^{-21}$ J |
| Charge on nucleus $Q_{NM}$ | $0.3 \times 10^{-11}$ Coul |
| Viscosity of cytoplasm | $\approx 10^{-3}$ (water) |
| Reynolds number | 462×0.4nm |
| Electron charge *q* (magnitude) | $1.6 \times 10^{-19}$ Coul |

This calculation, in Appendix B, follows a physical model, i.e. use of the laws of physics to model a given system. It assumes that life, on the most fundamental level, is, like all other known effects, subject to the fundamental laws of physics. But, of course, this should be subject to verification by laboratory observation. In fact such verifications exist (see [8b]).Thus the approach is supported by the usual principles of science.

The input ligand travels sequentially over two channels, from the CM-NPC and then from the NPC-DNA. The latter, containing the codons C,G,A,T/U, is the ultimate target, to be altered by the messenger protein. We consider these channels in turn.

### 3.4 Model assumptions for CM-NPC channel

Within the CM-NPC channel of cytoplasm, the proteins follow trajectories [6b,8a,b] (summarized in Appendix B) obeying Newtonian mechanics subject to the principle (2) of maximum Fisher information *I* about each site $x_0$ on the CM. It is found (Sec. 5.5) that a total instantaneous electric charge $Q_{NM} \approx +0.3 \times 10^{-11}$ Coul. exists on the nucleus, due to import from the cytoplasm of +charged NLSs *to a single NPC*. This is in reaction to the presence of the messenger protein (ligand) located radially away from the NPC at position $x_0$ on the CM, whose presence is signaled to the NPC either electrically or by a time gating effect (Sec. 7.3). The result will be found to be cell stability, to 2$^{nd}$ order in perturbations, and at a state of both low entropy and maximum order.

Each messenger protein has negative charge due to phosphorylation. It results that the protein travels by Coulomb attraction toward the single, positively charged target nuclear pore complex (NPC) on the nuclear membrane. It is also subject to drag and random diffusion by molecules of the cytoplasm.

We assume that the negative messenger protein is partially shielded from the positive NPC charge by negatively charged protein neighbors of the subject protein, but *is not shielded by the often assumed ionic charges* [8a]. The latter are very tiny and readily pass through the NPCs [8a] *before* the Debye-Huckel equilibrium distribution of charges is attained. Under this condition the hypothesis of maximum Fisher information implies a shielding length $l_D \approx 0.63 \mu m$ (reciprocal of central value $k_0 = 1.6 \times 10^6 m^{-1}$ found at Eq. (6) below). This defines a Coulomb force field of tens of millions of volts/meter. Indeed this has been observed [9].

### 3.5 NPCs independently and individually charged

A key proposal is that simultaneous with the charge-up of the target NPC *the charge on nearly all other NPCs is momentarily turned off* (Sec. 5.5). This has two benefits: (a) The resulting Coulomb force field maximally directs the protein toward its target NPC. (b) Using figures for a yeast NPC gives a net NPC charge of $Q_{NPC}$ = +0.29 × 10$^{-11}$ Coul. This agrees well with our previously assumed [8a] charge value $Q_{NM} \approx$ +0.3 × 10$^{-11}$ Coul. Note the alternative: Since a typical nucleus has thousands of NPCs, if these were all electrically charged simultaneously the resulting charge $Q_{NM}$ would be thousands of times too large.

Our working principle (a) *I* = maximum gives rise to an ordered, stable CM-NPC channel state that is (b) maximally far from thermodynamic equilibrium. We note in passing that Fisher information *I* is defined completely independently of the concept of entropy *H*.

It is interesting in this context to compare the growth processes of a crystal and a cell. As the crystal grows its order *R* (which is proportional to *I*, see Eq. (18)) grows, as does its entropy *H*. By comparison, as the cell grows its order *R* grows but its entropy level *H* decreases; the latter is in fact proven in Eq. (17). In fact, there, *H* decreases *because*, by principle (2), *I* increases. That *H* is maximum for the crystal but minimum for the cell emphasizes that crystals are dead whereas cells are alive. One must keep in mind, however, that this is merely for one channel of the cell, namely its cytoplasmic volume. Whether it holds as well within the nucleus is another question.

It is interesting in this regard that crystals are deterministic structures whereas cells have a random component, namely the randomly scattering molecules of the cytoplasm. Thus, somehow the randomness instantiates life. This might likewise be the case for instantiation of PI, since PI evidently requires a medium with high uncertainty, i.e. high Shannon entropy (or *H*), like Prigogine's dissipative structures.

### 4. MEAN FLUX *F* VALUE WITHIN CM-NPC CHANNEL

In general, flux *F* is the number of particles/area-time reaching a surface. At time $t_0$ = 0 each protein is located on the CM, at radial position $r = r_0$. Its final position in this CM-NPC channel is at radius $r = a$ of the NM. The mean protein flux at the NM is $F = \rho <v>$, where the mean velocity of each protein is $<v> = (r_0 - a)/(t_a - t_0) = (r_0 - a)/t_a$ since $t_0 = 0$ by hypothesis. Hence the mean $F = \rho(r_0 - a)/t_a$. This is a known, analytic function of the Debye-Huckel constant $k_0$ of the cytoplasm, as derived in [8a] and Appendices B and D. The approach ignores the particular fold/structure of individual proteins, seeking an average answer. Thus it ignores the possible selective transport of only certain proteins. However, the resulting predictions turn out to be in line with lab observations, and to predict highly stable (to 2$^{nd}$ order) structure of the CM-NPC channel.

### 4.1 Mean flux curve, smoothness effects
Fig. 1 shows the logarithm of the mean protein flux *F* traveling from the CM to NM surface. This curve has two key observable properties:

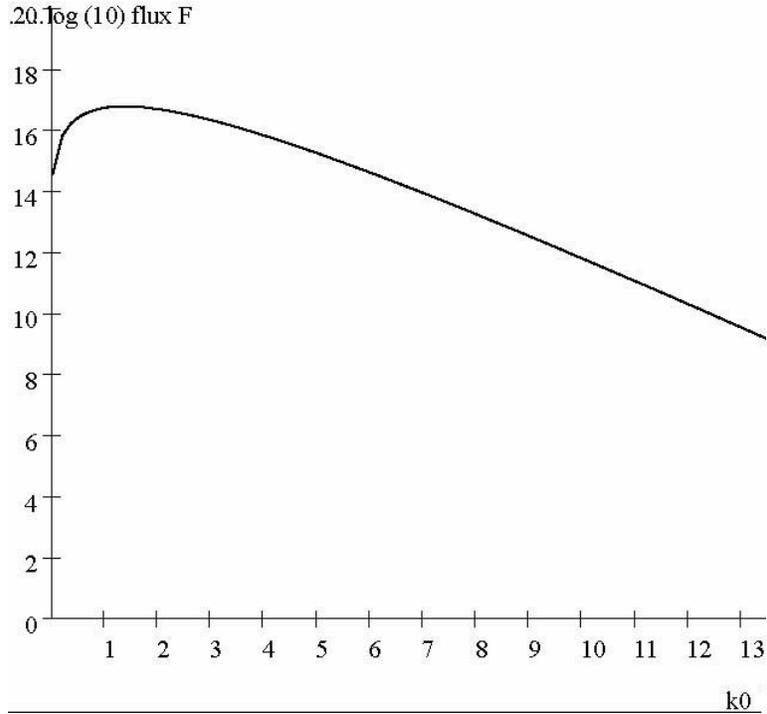

**Fig. 1. Mean flux *F* (proteins/area/ time) at the NM as a function of $k_0$ (× 10^6 m^-1)**

(1) It has a single region over which it is maximized. Thus there is a definite value $k_0 = k_{max}$ for which *log F* = max. and, therefore, *F* = max. ≡ $F_{max}$. The maximum is rather broad, of value

$$F_{max} = 10^{16.629} \approx 10^{16.6} \text{ proteins/(m}^2\text{s)}, \tag{5}$$

and occurring on about value

$$k_{max} = 1.6 \times 10^6 \text{ m}^{-1}. \tag{6}$$

The value (5) for $F_{max}$ compares well with laboratory measurements, which indicate that the NM is capable of accommodating a maximum flux rate of $\approx 10^{16}$ proteins/m²-s. This is shown next.

With NM radius *a* = 3 μm, total surface area *A* = 3.14×10⁻¹⁰m². Each NPC can accommodate the passage of about 10³ proteins/s [8a], and there are about *3000* NPCs in a mammalian cell. This allows a total flux *F* = (*3000×10³* proteins/s) / (*3.14×10⁻¹⁰*m²) ≈ *10¹⁶* proteins/m²-s.

This is less than an order of magnitude departure from the derived upper limit (5) of 10^16.6 -- fairly good agreement. The curve in Fig. 1 also shows a strong decrease (by orders of magnitude) once $k_0$ is greater than roughly 4.0 ×10⁶ m⁻¹.

By definition of the maximum,

$$\frac{dF_{max}}{dk_0} = 0 \text{ } at \text{ } k_0 = k_{max}. \tag{7}$$

However, the second derivative at $k_0 = k_{max}$ is also of interest. The maximum $F_{max}$ is attained over a relatively broad region of $k_0$ values. This is apparent by zooming in on Fig. 1 to emphasize the region near its maximum $F_{max}$; as shown in Fig. 2.

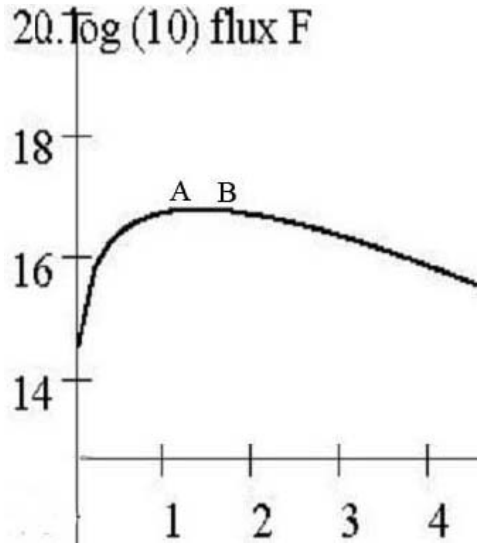

**Fig. 2. Close-up of region AB containing the maximum $F_{max}$. The abscissae are values $k_0$ (x 10^6 m^-1) as in Fig. 1. The curve is close to flat over span AB, consisting of about 12 plotted points.**

This shows that the region AB of values $k_0 = (1.1, 1.7) \times 10^6 \, m^{-1}$ containing the maximum is close to flat. In fact, it *is flat – at value $log_{10} F_{max} \approx 16.63$* -- within the thickness of the curve. This is *over about 12 plotted points comprising the region AB centered on the maximum value $F_{max}$.* (Pixel width may be observed as the individual step widths in the long tail of the curve to the right.) *Thus its 2$^{nd}$ difference, and consequently 2$^{nd}$ derivative, about the* central point $k_0 = 1.6 \times 10^6 m^{-1}$ is close to zero,

$$\frac{d^2 F_{max}}{dk_0^2} \approx 0 \text{ at } k_0 = k_{max} = 1.6 \times 10^6 m^{-1}. \qquad (8)$$

Effects (7) and (8) have important consequences of stability, as discussed next.

### 4.2 Resulting stability of flux to 2$^{nd}$ order

The measured ratio of the plotted curve thickness to the plotted vertical distance between values 16 and 18 of $log_{10} F$ is about 1/34. Thus the curve thickness represents a change

$$\Delta \, log_{10} F = (1/34)(18-16) = 0.0588 \qquad (9)$$

in the log of the flux ($log_{10}$ proteins/m$^2$-s). Now the relative change $dF/F$ in flux $F$ obeys generally

$$dF/F = d(lnF) = d[(ln10) log_{10} F] = 2.303 d[log_{10} F]. \qquad (10)$$

Therefore by (10), since the changes are small, the change (9) due to thickness of the curve translates into a relative uncertainty $dF/F = 2.303(0.0588) = 0.1354$, or *13%*; relatively small. This is for the region AB on the curve, where the abscissa values $k_0$ change from value 1.1 to 1.7 (times $10^6$ m$^{-1}$). This is a relative change in $k_0$ of about $(1.7 - 1.1)/1.1 = 0.545$ or *54%; relatively large*.

Hence, a large relative change in $k_0$ of *54%* gives rise to a small relative change of *13%* in the flux $F$. What changes in the cell parameters describe such a change scenario? The Debye-Huckel parameter obeys Eq. (4). By taking differentials of both sides of it, the *54%* relative change in $k_0$ translates into either a *27%* change in $\rho$, $\varepsilon$ or $T$, or a *54%* change in $q$. These changes represent significantly large perturbations. Thus, the flux $F$ in the vicinity AB of its maximum obeys relative insensitivity (of order *13%*) to perturbations in these cell parameters. This has very important ramifications of stability for the system.

By Taylor series, we may generally represent the resulting perturbation in $F_{max}$ as

$$\delta F_{max} = \frac{dF_{max}}{dk_{max}} \delta k_{max} + 2^{-1} \frac{d^2 F_{max}}{dk_{max}^2} \delta k_{max}^2 + 6^{-1} \frac{d^3 F_{max}}{dk_{max}^3} \delta k_{max}^3 + \ldots \qquad (11)$$

By Eq. (7) the first right-hand term is zero. Therefore for small changes $\delta k_{max}$ the main contribution to $\delta F_{max}$ is from the 2$^{nd}$-order term in (11). But we found that $\delta F_{max} / F_{max} \approx 13\%$, small. Therefore, assuming a smooth curve over region AB with a single maximum located about halfway, the 2$^{nd}$-order term is likewise small.

Consequently, $F_{max}$ is approximately invariant through 2$^{nd}$ order in *all system parameters that define* $k_{max}$. As we saw above, these include perturbations $\delta T$ in the temperature, $\delta \rho$ in the mass density of proteins, $\delta q$ in their electric charge and $\delta \varepsilon$ the permittivity of the cytoplasmic medium. Hence *the cell state is stable through 2nd order in all such environmental changes.* Although such stability may not be sufficient for generating life, it is certainly helpful. Note that this result is only for the limited channel under consideration. We have yet to analyze the NPC-DNA channel wherein the genome actually exists.

### 4.3 Effective 'emergence' of stability

The flatness of the flux curve over region AB in Fig. 2 is suggestive of the flat extremum that characterizes a double-well potential as the two potential sources approach one another. A flat extremum emerges, for example, as the sum of a Gaussian well potential - exp(-x²) and its displaced version - exp(-(x-√2)²). Double-well potentials are known to generate 'emergent' physical phenomena such as superconductivity. Hence the flat region AB of flux in Fig. 2 suggests the emergence of a stable life phenomenon. (Again, this is only for the CM-NPC channel under consideration here; see Sec. 9 for the NPC-DNA channel.)

As the development has shown, such emergence ultimately follows from the premise (2) of maximum Fisher information. See the paragraphs following Eq. (2) for supporting information.

## 5. CONSEQUENCES

### 5.1 Information *I* at NM

These effects (8), (9) on flux affect information *I* as well. First, since *I = max* by Eq. (2) it follows that if the system is perturbed the first-order perturbation of the information obeys δ*I = 0.* Next, the Fisher information *I* about protein position at the NM *is proportional to F*, through Eq. (3). Then the 2$^{nd}$-order effect Eq. (8) implies that δ$^2$*I = 0*. In summary, the information obeys both

$$\delta I = 0 \text{ and } \delta^2 I \approx 0. \quad (12)$$

at a Debye-Huckel shielding parameter value $k_0 = 1.6 \times 10^6 m^{-1}$. This describes *stability to both first and second order* environmental perturbations to the cell operating at point $F_{max}$ on the curve Fig. 2. Second-order stability rarely occurs in physical systems, and might account for the impressive stability of living systems

In contrast with physical systems, living systems exist by virtue of 4 billion years of natural selection, where the fittest survive. At each generation there are tiny, random changes in the DNA. By natural evolution only those changes that are advantageous give fitness advantages, and so only these specimens ultimately survive and dominate. Over 4 billion years this can make for very large changes in population makeup. Basically, time acts as a slowly acting filter of the fit from the unfit, and one component of fitness is stability to environmental perturbation.

It might clarify the picture to understand that the information *I* that is so stabilized is not a property of the environment but, rather, of the CM-NPC channel. To be precise, *I is both maximized and stable* to random perturbations of protein density $\rho$, the electric charge *q* on each protein, the dielectric constant $\varepsilon$ and temperature *T*. These are all properties of the cell cytoplasm and its proteins. However, these properties can be randomly perturbed by environmental influences such as a change of temperature, sudden influx of proteins, influx of ions of a different dielectric constant, etc. The analysis above predicts, at Eq. (12) that, nevertheless, the cytoplasmic level of information *I* will be stable to these environmental influences.

There also are implications as to the number of different types of protein making up a pathway, the resulting Debye-Huckel length, the resulting total charge on an NPC and the charge on the nucleus per se. These are taken up next.

### 5.2 Number of protein types per pathway

Since the number of proteins $n \approx k_0^2 \times 10^{-12}$ m$^2$ (by Appendix S2 of [8a]) the maximum flux value is accomplished by a pathway containing about $n = (1.6 \times 10^6)^2 \times 10^{-12} \approx 2$ types of protein. But given the abovementioned flatness of the uncertainties in the inputs to the calculation a more realistic figure is a range $n = 2 \pm 1$ protein, or $n = 1,2,3$. This accounts for the predominance of the RAF, RAS, MEK, etc., pathways of cell function.

### 5.3 Debye-Huckel length

The value of $k_{max} = 1.6 \times 10^6$ m$^{-1}$ gives a corresponding Debye-Huckel length

$$l_0 = 1/k_{max} = 0.63 \mu m. \qquad (13)$$

Note that this assumes Coulomb screening only due to neighboring messenger proteins, in particular *not inorganic ions* as previously discussed in Sec. 3.4.

### 5.4 Resulting NPC charge

Adopting the development in [10], and using result (11) for the Debye-Huckel length $l_0$, the charge on an NPC obeys

$$Q_{NPC} = (8 \times 13)/(41962 \text{ nm}^3) \times (2\pi/3)(0.63 \times 10^3 \text{ nm})^3 \times 14q$$
$$\approx 0.29 \times 10^{-11} \text{Coul.} \qquad (14)$$

Note that $q$ = electronic charge in Table 1.

### 5.5 Resulting Charge on nucleus

There is evidence [10-15] that NPCs act *independently* to attract specific NLS sequences and messenger proteins. We propose that each NPC turns 'on' the computed positive charge value (12) of $Q_{NPC} = 0.29 \times 10^{-11}$ Coul., while the charge on *all (or nearly all) other NPCs are turned 'off.'* Then the proteins travel in near *straight-line* fashion (save for minor diffusion in the cytoplasm) from CM to target NPC. This minimizes the traversal distance, and hence traversal time $t_a$, from CM-NPC.

The NPC attains the charge in the usual way, by absorbing an appropriate NLS from the cytoplasm. After this protein enters the NPC a different NPC turns "on" so as to Coulomb-attract its target protein, with all other NPCs turned 'off'; etc. In this way the net nuclear charge at any time is

$$Q_{NM} = Q_{NPC} = 0.29 \times 10^{-11} \text{Coul.} \qquad (15)$$

This is quite close to the value $0.3 \times 10^{-11}$ Coul previously assumed [8a] for the nuclear charge $Q_{NM}$. It is also confirmed by experimental work of Tyner et al. [9]. These authors report position-dependent electric field $E$ values in the cytoplasm – measured by use of a novel nano-voltmeter -- that are consistent with this $Q_{NM}$ value. These are values of field strength $E \sim$ tens of millions of V/m. Thus, works [8a] and [9] support the cell model proposed here, including the property that one NPC at a time admits proteins; see Sec. 8 for further laboratory verifications.

### 5.6 Protein positional uncertainty at NPC

The proposed model would only work if the uncertainty in position of proteins at the NM is significantly less than the functional opening. The latter is about *9* nm, although openings of *39* nm have been proposed. Use of the value $F_{max}$ from Eq. (5) in Eq. (3) and the parameter values listed below it give a numerical value of $I=I_{max}=2.83\times 10^4$ $\mu$m$^{-2}$. Then by the second Eq. (1),

$$e_{min} = 1/\sqrt{I_{max}} = 5.94 nm. \qquad (16)$$

This is roughly half the NPC opening so that, for the Gaussian pdf assumed, about *95%* of all incident proteins will enter their targeted NPC (or of course much higher for the *39* nm-opening proposed).

## 6. MAXIMUM ORDER AS A SOURCE OF NON-EQUILIBRIUM

On their travel through the CM-NPC information channel, the messenger proteins randomly encounter molecules of the cytoplasm and, so, randomly diffuse from their otherwise straight-line Coulomb-enforced trajectories toward corresponding NPCs. The molecular interactions occur independently, so that the central limit theorem [5] predicts noise values $x$ at the NPC that obey a normal law $p(x)$. Let this have variance $\sigma^2$.

### 6.1 Dependence of entropy *H* upon information

For the normal law *p(x)*, information $I = 1/\sigma^2$ [1] and entropy $H = \frac{1}{2} + \ln[(\sqrt{2\pi})\sigma]$. Eliminating the common parameter $\sigma$ between these expressions gives

$$H = 2^{-1}[1+\ln(2\pi) - \ln I]. \qquad (17)$$

This is a relation connecting the Shannon entropy with the Fisher *structural* information. Again, this is only descriptive of the CM-NPC channel, i.e. *exterior to* the nucleus. Within this CM-NPC channel it gives a unique value of the entropy *H* for each value of *I*. It shows, e.g., that *H* is low when *I* is high. This makes intuitive sense as well, verifying that disorder *H* is low when order *I* is high. However, it should be kept in mind that *I* is the Fisher information. This is mathematically and physically well defined (Appendix A). In particular it is not "prescriptive information" PI of genetic selection (GS) theory. Since also Eq. (17) holds exterior to the nuclear channel containing the genomic DNA information, (17) in no way infers a relationship between entropy *H* and *PI*.

By Eq. (17), the entropy *H* of the CM-NPC channel drops ever more below the thermodynamic equilibrium state $H_{max}$ as information *I* increases. This gives rise to two beneficial effects:

### 6.2 Entropy is minimized

Since, by Eq. (2), *I* = max., by (17) the CM-NPC channel will operate at a level of *H* = min. By the connection *S = kH* Boltzmann entropy *S* and Shannon entropy *H*, this likewise means a small entropy *S* as well. There is only a small degree of randomness in fluctuations *x*. The system has increased order. This is indicated, as well, by *S* being, now, *maximally far below* its equilibrium value $S_{max}$, the latter defining death in fact.

It is interesting to interpret these two results further. Achieving $H_{max}$ is equivalent to knowledge of maximum uncertainty. Likewise, achieving the corresponding $S_{max}$ describes a scenario where all energy within the channel is waste heat, so that there is no thermodynamic free energy left to sustain motion or build life, that is, heat death is reached.

Moreover, the channel will be stable to up to second-order perturbation, as shown by Eq. (12).

### 6.3 Order is maximized and stable: Prigogine's goal

The measure of order *R* appropriate to a system with a continuous coordinate *x,* as here, is [16]

$$R = (1/8)L^2 I. \qquad (18)$$

This measure was derived from the *defining property* that it decreases (or stays constant) when the system loses structural detail, by a random process called "coarse graining". For example, the intrusion of a pathogen, such as a virus, into a system is random to the system (though perhaps not to the pathogen, which may be carrying through a growth program of its own). Thus it amounts to a random perturbation of the system. Eq. (18) has the property that it will go down in this case (see below for further on coarse graining).

*6.3.1 Order R is not dependent on Shannon entropy*

So as not to lead to confusion, it should be emphasized that the measure (18) of order depends on the Fisher, not the Shannon, entropy. Thus it has nothing to do with 'bits' of information in the system either before or after the coarse graining event. Instead, it depends upon the Fisher information *I,* and hence, by Eq. (1), the level of *local structural detail* in the system probability law *p(x)*. By Eq. (1) this, in turn, measures its total amount of gradient content $(dp/dx)^2$. Thus, it is the degree to which *p(x)* exhibits fast *up-down structure* that contributes to its value and hence, by Eq. (18), to the level of the order *R* . Thus, the latter is effectively a measure of the amount of local 'structure' in the system *p(x)*. The mathematics of the derivation equates this to the order. By comparison, the Shannon entropy is a 'global measure,' insensitive to local structural detail.

*6.3.2. Order R measures complexity as well as degree of order*

*R* also measures the degree of 'complexity' in the system. Consider, e.g., a two-dimensional probability law *p(x,y)* containing *n* sinusoidal ripples in each direction over a field *0 ≤ x ≤ 1, 0 ≤ y ≤ 1*. Thus there are $n^2$ total ripples in the field. See Fig. 3.

**Fig. 3. Sinusoidal PDF with *n* = 5 ripples in each of 2 dimensions (x,y). Note the visually high level of complexity that accompanies the high level of order *R*.**

This law obeys $p(x,y) = 4 \sin^2(n\pi x) \sin^2(n\pi y)$. Using this in definition (18) gives the value of the order as exactly $R = 2\pi^2 n^2$. But we found above that the total number of ripples in the scene is exactly $n^2$. Or in this case it is 5 x 5 = 25, as can be easily counted in Fig. 3. The ripples are the individual 'details' of the scene. Thus, from the above, the order *R* is $2\pi^2$ times *the number of details in the scene*. Or, it is *proportional to* the number of such details.

This 'number of details' also agrees with *a measure of complexity* (rather than 'order') due to Kolmogoroff and Chaitin. This is called the Kolmogoroff-Chaitin (K-C) complexity [17]. The K-C complexity is likewise proportional to $n^2$ in application to the Dijkstra [18] routing algorithm. The K-C measure also equals the total number *n* of statements in a computer program, the total number *n* of switches within a network; or the shortest description of a string in some fixed universal language.

*6.3.3. Effect of coarse graining is to decrease order R*

Thus, due to random coarse graining events, as time increases system order *R* (and, by (18), information *I*) tends to *decrease*. (This ignores order-building effects due to energy inputs, described next.) The time increase defines an *arrow of time* (See Appendix A). In particular, for coarse grained living systems it is the oft-discussed *biological* arrow of time. It should be emphasized, however, that a cell whose proteins obey principle (2) of maximum *I* (or *R*) *is not* undergoing coarse graining. To the contrary, it is building information and order by the constructive use of energy inputs from the environment (Sec. 6.4). These act as 'functional information' inputs.

*6.3.4 Attaining the converse of Prigogine's goal*

In Eq. (18), length *L* is that of the largest straight line through the system. Then for the cell under study *L = 2a =10* μm. Since, again, by Eq. (2) *I = max*, Eq. (18) shows that the CM-NPC channel will have *maximum order R*. Thus, the headings **6.2** and **6.3** together literally state that, for this channel, the demand for maximum order *R* becomes a source of non-equilibrium (minimum entropy *H*) for the system. I. Prigogine's goal [19] was to show the converse, that "non-equilibrium (*H = min.*) becomes a source of order" *R (R = max.).* In fact this directly follows from Eq. (17), since $R \propto I$ by (18).

We also saw, below Eq. (10), that *despite* such operation at far from thermodynamic equilibrium, the maximized order level (18) *is stable to second order* environmental perturbations (in particular, over region AB in Fig. 2). So, not only is the order maximal but it is stable as well. Comparison with the corresponding growth of a crystal, in Sec. 3.5, are apt here. Thus, if it were not for random components in the otherwise regularities of a crystal, its level of entropy would be maximum, its level of order minimum (the opposite of the situation in the living cell).

Finally, in Sec. 4.3, we saw that this $2^{nd}$-order stability is suggestive of the flat extremum characterizing a double-well potential as the two potential sources approach one another. Double-well potentials are known to generate 'emergent' physical phenomena such as superconductivity. Here the emergent phenomenon is stable life. But again, this has limited validity since we are only addressing the CM-NPC channel, which does not contain the genomic DNA information.

**6.4 Key role of environmental energy inputs**

Our premise is (2), that information is maximized; then by (18) so, likewise, is the order *R*. On the other hand, Eq. (18) for *R* was derived on the premise that *R must decrease* after each coarse-graining event such as a cancer cell input. However, it should be kept in mind that the cancer follows its own 'program' of growth; as do other cell invaders such as the Plasmodium parasite. And in so succeeding these can ultimately kill their host, whose levels of order go to zero. How does a cell cope with such a deteriorating situation, and instead gain order?

To provide an answer requires introduction of the concept of 'coarse graining' [16]. In general a system is coarse grained when its finest structural details are lost due to some physical process. An example is replacing fine-grained photographic film by, literally, 'coarse grained' film. There is always a visually obvious loss of resolution (and, so, coarse graining).

The answer to the question of the $2^{nd}$ preceding paragraph is that cells utilize imported environmental energy [20], such as photons to promote photosynthesis, and internal energy sources, e.g., ATP and GTP. These are used to create the binding energy required to form deterministic cell structure, and order. Until the cell approaches senescence, these counter the deteriorating effects of 'lossy' coarse graining events such as influx of cancer. For example, ATP and GTP power the import of positive NLS charges into each target NPC (Sec. 3.4), so as to form a sufficiently large electric field *E* to pull a maximum level of messenger protein flux *F* into their target NPCs. Then, since flux $F \propto I$ = max. [by Eqs. (2) and (3)], and since the rms error of location $e_{min} = 1/\sqrt{I}$ [Eq. (1)] at each NPC, the error must be minimized. As we saw (Sec. 5.6) it results that 95% of proteins successfully enter their target NPCs.

Another use of the ATP and GTP energy resources is to power the chemical reactions necessary to move each required ligand along the DNA spiral to its target location (Sec. 9).

**7. TRAJECTORIES IN THE CM-NPC CHANNEL**

**7.1 Charge effects**

Suppose that a ligand arrives at a specific position $x_0$ on the inner surface of the CM? Which NPC on the NM should be its target? By our minimal time requirement (above Sec. 3.1) the target NPC should be located radially inward from the ligand position $x_0$. This is by the following reasoning.

We have *assumed* a ligand that travels from CM to NPC to DNA. Therefore, here we consider what would make the negatively charged ligand enter the NPC. Elementary electrostatics suggests that this is accomplished if the NPC can somehow acquire a positive electric charge (15). Recent research [10] indicates it does so by importing NLS molecules from the cytoplasm. This charge on the NPC is strong enough that, aside from minor excursions due to diffusion by the cytosol, the ligand travels *radially inward* toward that NPC. Of course such a radial path is also the shortest possible distance from the NPC to any possible ligand position on the CM. Hence the ligand travels to the NPC in *the shortest possible time*, which we called $t_a$. For the given spherical cell model, $t_a$ = 0.016 s [8a]. Also, because of the minimal random spread (16) in position on the NM due to diffusion, the probability that the target ligand will enter the NPC is about 95% (as derived below Eq. (16)). Also, to discriminate against other, *non-radially* located ligands, the NPC maintains its charge for the limited time $t_a$. Such non-radially positioned NPCs would require more time to get to it. This is why the targeted NPC should be the one located radially inward from ligand position $x_0$.

In this manner, all ligands take time $t_a$ to travel radially inward to their corresponding NPCs. This constant time $t_a$ is built into the operation of each NPC -- It maintains its Coulomb charge, and keeps its central plug open, only during the time $t_a$ ± small tolerance time.

**7.2 Use of possible signaling mechanism**

First, how does a given NPC know *when* to turn 'on' its charge so as to acquire the target ligand? It would suffice, e.g., if the presence of the ligand at $x_0$ were signaled to the nucleus by a nanosecond pulsed electric field originating on the CM [21].

**7.3 Alternative use of time gating mechanism**

But, what if such a pulsed electric signal does not exist? Now how can the NPC acquire its target ligand? First, how does a given NPC know when to turn 'on' its charge so as to acquire the target ligand? One possibility is that the NPC utilizes a time gating procedure. As above, it turns on its charge for the fixed time $t_a$. If that type ligand was already present radially away on the CM when the NPC turned on its charge, then that ligand and only that ligand can enter the NPC. However, if that ligand arrived on the CM at some time after the "on" phase of the NPC started, it wouldn't get there in time. The NPC would be plugged, and this would be a missed signaling opportunity. Meanwhile, there is also a chance that a nearby ligand in the cytoplasm from some other CM position could be pulled into this NPC. This would constitute an erroneous ligand input. However, the chance of this happening can be minimized if, sequentially in time, "on" state NPCs are *widely separated* angularly around the NM. (For example, consecutively open NPCs are located at least 90° apart.) Then the chance of such an errant ligand being close enough to the target NPC to be pulled in would be minimal. In this way, there is effectively minimal probability that an oblique-motion (non-radial) ligand will enter the subject NPC. So the NPC plays a "waiting game," sequentially over time intervals $t_a$, until it gets the ligand it 'wants.'

This argument can be extended to non-spherical cells and nuclei as well. All ligands entering NPCs will always originate on the CM the minimum distance from its target NPC. In this way, by selective "time gating," a given NPC receives only ligands that are biologically "conjugate" to it, i.e. originate across the cytoplasm at a single corresponding point $x_0$ on the CM. But also, we have to consider that there are different types of ligands, and perhaps the NPC has incompatibility with the particular one that entered. That would require some subsequent rejection of the ligand within either the NPC or its chromatin region. These considerations are outside the scope of this paper.

**8. LABORATORY VERIFICATIONS**

The model gives results that have been experimentally verified [8a,8b] *Note*: In [8b] see in particular the next-to-last section "Supporting Evidence: Summary":

(i) Very high intracellular electric field strengths, typically tens of millions of volts/meter.
(ii) A central role for negative charges, added to proteins by phosphorylation, in promoting their Coulomb force-dominated motion toward the positively charged nucleus;
(iii) The dominance of protein pathways consisting of from 1-4 proteins, e.g. the RAF, RAS and MEK pathways;
(iv) A fast response (2,800 proteins/*ms*) of cells to sudden trauma such as wounds;
(v) A *4nm* size for the EGFR protein, which has been observed to be of about size *3nm.*

## 9. INFORMATION FORMS FOR BOTH CM-NPC AND NPC-DNA CHANNELS

In the foregoing we analyzed one information channel of the cell: the information carried by a protein in moving from the CM to an NPC on the NM through a channel of *continuous* (in coordinate *x*) cytoplasm. We found that Fisher information *I* characterizes the order *R* in this cytoplasm channel, via Eq. (18). But this is only half the total information channel. The rest consists of the protein that entered the NPC continuing on to a target appropriate DNA sequence; there it alters one of its codons. Such a sequence is, of course, discrete.

### 9.1 The NPC-DNA channel
A DNA molecule consists of codons C,A,T,G, etc. in some discrete sequence that generates PI. However, Fisher information *I* is not definable for discrete sequences. Therefore Eq. (18) for *R* cannot be used to define its level of order. Then how can the order in *this* process be characterized?

### 9.2 Model assumptions for NPC-DNA channel
The foregoing results followed from the principle that the Fisher *I* about ligand position on the CM for proteins traveling through the cell cytoplasm obeys *I* = maximum. But, what principle is followed inside the nucleus? A well-defined statistical answer is to seek the maximum probable DNA binding site.

### 9.3 Kullback-Liebler measure
Many workers (e.g., [22], [23]) characterize the order in DNA sequences by their Kullback-Liebler (or "cross-") entropy

$$H_{KL}(P_n, Q_n) = \sum_{n=1}^{N} P_n \ln\left(\frac{P_n}{Q_n}\right). \tag{19}$$

Here the probabilities $P_n$ define the given system and the $Q_n$ are "reference" probabilities prescribed by the user.

One use [22] of K-L form (19) is as an enzyme function predictor. There (19) is a distance measure for defining the "enzyme commission number" EC. This classifies an enzyme based on the chemical reactions it catalyzes. A second use of KL-entropy (19) is in identifying a DNA sequence's function as either a *regulatory protein* or a *restriction enzyme;* see, e.g., Stormo [23]. He gives

$$H_i(P, Q) = \sum_b P_{bi} \log_2\left(\frac{P_{bi}}{Q_b}\right) \tag{20}$$

as the *average binding energy* at a given ligand binding site *i* on the DNA chain. (See further on this following Eq. (21).) Here $P_{bi}$ is the probability of codon base *b* occurring at site *i*, and $Q_b$ is the total probability of base *b* occurring over the entire genome. The binding energy associated with a specific substrate-enzyme interaction significantly lowers the Gibbs *free energy* change required for the reaction.

For Eq. (20) to represent the K-L entropy of the entire DNA molecule it should be summed over spatial positions *i* as well. Assume that the *total* entropy $H(P, Q)$ is simply the sum of energies $H_i$ over the individual binding sites *i*. Then from (20)

$$H(P, Q) = \sum_{bi} P_{bi} \log_2\left(\frac{P_{bi}}{Q_b}\right). \tag{21}$$

(This assumption of additivity of energy is a limitation of the approach.) Note that *H* and *S* are proportional; see Sec. 2.5. It results that (21) represents total energy (at constant *T*) as well. But note that $\sum_i P_{bi} = Q_b$ so that, expanding out the $log_2$ quotient in (21) gives a term $\sum_{bi} P_{bi} log_2(Q_b) = \sum_b Q_b log_2(Q_b)$. The latter is the total mean energy of the codons, *irrespective of* their locations. Interestingly, some of the most intricate biological instructions require minimal energy expenditure. Since in (21) *this subtracts from the total energy in all codons b at all sites i*, the result is the total binding energy *specific to the sites,* i.e. the total binding energy as was desired.

Eq. (21) is now in the form Eq. (19) of a K-L entropy, with $Q_b$ the reference function. By elementary thermodynamic considerations, the probabilities $P_{bi}, Q_b$ that mathematically *maximize* binding energy $H(P,Q)$ are found [23] to form *maximum probable* binding sites. Hence, a principle

$$H(P,Q) = \text{maximum} \qquad (22)$$

is obeyed by the placement of codons in the DNA. But of course DNA sequences are dynamic, not static. This dynamic nature is usually attributed to random mutations. Mutations either add, delete, or alter sequences of base pairs. Rarely, a mutation will benefit the fitness of the organism. This mutation then has a greater chance of being passed on to its offspring and perpetuated. This is a randomly driven creation of new information.

By comparison, PI proponents state that natural selection is only eliminative, not creative. This would seem to say that a beneficial mutation will not be passed on to its offspring (denying one of the basic tenets of natural selection).

Note also that proponents of PI question a connection between this entropy (22), or its approximating Fisher information (23), or even its level of order, to genome function. In reply, we can state that the pillars of scientific theory rest upon a foundation of laboratory verifications. And there have been many of these, as summarized in Sec. 8.

### 9.4 Approximation by a Fisher measure

Finally, it is interesting to consider whether this discrete measure, for use on the discrete DNA channel, could be consistent with the preceding use of Fisher information *I* for the *continuous* CM-NPC cytoplasm channel. Suppose that the local codon probabilities $P_{bi}$ do not strongly differ from the *global* value $Q_b$ independent of position, i.e. $P_{bi} = Q_b - \Delta P_{bi}$ where all $\Delta P_{bi}$ are small. Then from (21) the total binding energy over codons is

$$H(P,Q) = \lim_{\Delta P_{bi} \to 0} \sum_{bi} P_{bi} \log_2 \left( \frac{P_{bi}}{P_{bi} + \Delta P_{bi}} \right) \to 2^{-1} \sum_b I_b = 2^{-1} I_{DNA}. \qquad (23)$$

This particular limiting answer is derived in Appendix C. The general derivation is in [6a], pgs. 37-38, and [24]. Hence, in this case the total mean binding energy $H(P,Q)$ is a discrete approximation to ½ the Fisher information $I_{DNA}$ for the DNA channel. Then, serendipitously, by principle (22) of maximum probable (or MAP) [25] binding sites, principle (2) of maximum Fisher $I_{DNA}$ for the DNA channel actually *derives* here.

A limitation of the approach is that the above assumption of independent binding site energies is mathematically equivalent to assuming that the codons act as *independent* sources of information.

In summary, under these conditions the single extremum principle (2) of maximum Fisher information acts to effect and describe the overall cell development channel (CM-NPC-DNA). See Table 2. With time increasing to the right, the chief events of the particular channel (bottom row) are shown that result from operational effects (top row) upon them:

**Table 2: Chief events of the CM-NPC-DNA channel (bottom row), and their causative effects (top row)**

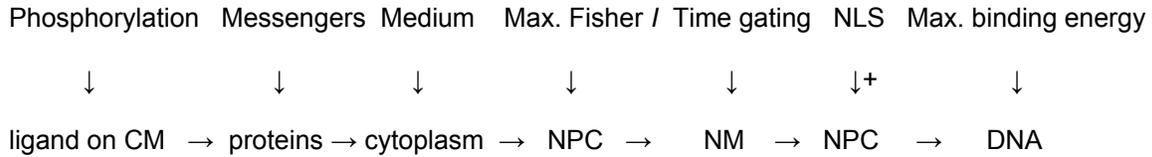

Use in this paper of the concept of Fisher information to quantify the development of life processes has had further success in quantifying the development of entire ecologies [26].

## 10. DISCUSSION

We have reviewed the information basis for the following aspects of normal cell growth:

(1) Natural selection has produced ligand-carrying proteins that travel from CM-NM in minimal time (Sec. 3). This permits, e.g., quick functional cellular response to sudden environmental trauma such as wounds or foreign bodies.

(2) As a consequence, cells obey maximum *structural* information $I$, subject to constraints of limited energy, passage through membrane barriers, etc. In this application of the principle, this resulted in minimal time $t_a$ of passage of messenger proteins from the CM to the NM. It also resulted in beneficial stability of the system to approximately $2^{nd}$ order (Secs. 4.2, 4.3, 5.1), a rare case in nature. Of course the cell did not *know* to evolve in this direction. It was simply following the principle $I$ = *max.* of having maximum up-down structure in $p(x)$. In this case, it is a $p(x)$ that has optimally high slopes because it is a Gaussian function with small variance $\sigma^2$. That this resulted in an advantage of survival is mere happenstance. Again, that is the nature of natural selection.

But of course such optimal function is not limited to biotic systems. Inanimate systems, for a much longer period of time, have been known to obey extremum principles: minimum time of travel (for a light ray), extreme Lagrangian action, EPI, etc. Even natural selection per se has been hypothesized as dictating the evolution of black holes into universes [L. Smolin, "Baby Universes" book (Sec. 3.2) at the NM and maximum average flux $F$ (Sec. 4) in the cytoplasm. This results in stability, *approximately through second-order*, in both their protein flux rates and spatial information levels at the NM (Secs. 4.1-4.2).

(3) During protein acquisition by the NPCs each NPC individually acquires a positive charge via NLS acquisition, while nearly all other NPCs remain uncharged (Secs.3.5, 5.5). This results in efficient protein acquisition by allowing each such negatively charged molecule to be directly guided toward its (positive) target NPC on the NM. Also, using newly understood NPC biology [5], we derive the experimentally observed charge $Q_{NM} \approx +0.3 \times 10^{-11}$ Coul that was assumed in past calculations of Fisher-based cell dynamics,.

(4) In the absence of a synchronizing field from the CM, each NPC utilizes a time gating procedure (Sec. 7.3) to select its target protein from among all others in the cytoplasm.

(5) The principle Fisher $I$ = maximum is obeyed by proteins moving within the cytoplasm (Sec. 3.2). It leads to a CM-NPC channel that has maximum order $R$ (Sec. 6.3) and that is both highly stable (to $2^{nd}$ order environmental perturbations; Secs. 4.2, 5.1) and maximally far from thermodynamic equilibrium (Sec. 6.2). This leads to the 'emergence' of life (Sec. 6.3).

This begs some questions (of a reviewer):

(a) *Why* do such proteins manifest maximum Fisher Information? This is a deep question. Ref. [2] shows that all textbook physics, i.e. of inanimate systems, obeys the principle. Why should this be? The view taken in [2] is that the inanimate Universe generally favors the acquisition of accurate information (high $I$). Regarding living systems, it seems to be true only of eukaryotes. Prokaryotes and cancer cells, in fact, appear to obey a principle $I$ = *min*. (again, subject to the constraints of the problem) [6b].

(b) Does just *any* polyamino acid stochastic ensemble manifest maximum Fisher Information? Or just functional proteins? How did poly-amino acid polymers get to be true proteins, and how did they get to be functional (only one out of $10^{77}$ stochastic ensembles has any protein function)? Are functional proteins exclusively determined by physicodynamics? Or are they determined by polycodon sequencing/programming and by ribosomal algorithmic processing, both of which are formally determined and controlled?

In partial answer, there is no one maximum Fisher information answer. The constraints imposed by the and molecular makeup of the system affect the maximization. Each of the preceding questions would probably have a different set of constraints. At any rate, these are likewise very good questions, worthy of consideration for future research.

By the way, it is the various sets of constraints that lend variety to the answers: In response to a reviewer, the information *I* approach does not give boring, redundant, "low-informational" (quite the opposite) and stable-ordered systems. Each case gives a different answer, just as each physical scenario gives a different law (whether Newton's, Maxwell's, or Schrodinger's).

(6) Within the NPC-DNA channel a principle of maximum binding energy [23] (or Kullback-Liebler entropy) is obeyed (Sec. 9.3). According to PI, the linkage between maximum binding energy and biofunction arises from the PI of codon sequencing that prescribes amino acid sequencing that determines tertiary structure of proteins, their hydrophobicities, charges, grooves, pockets, etc. Finally, under small-change conditions this goes over into a principle of Fisher *I* = maximum (Sec. 9.4); then the entire CM-NPC-DNA channel obeys the Fisher maximization principle Eq. (2). This principle is, on the face of it, descriptive rather than determinative.

However looking deeper into assumptions behind its steps often leads to determinative conclusions as well. For example, it (in the form of EPI statement [2] that $I - J = min.$, or equivalently $I \approx J$) predicts the Schrodinger wave equation. The latter physical law is then realized to have resulted from the deeper statement that particles move in such a manner as *to convey maximal positional information* $I \approx J = max.$ to the observer. This is beyond mere description. Likewise, the use of descriptive information DI might imply a level of prescriptive information PI, depending upon case. But this is mere speculation.

(7) Many of these effects have been verified experimentally, as discussed in Secs. 4.1 and 8.

These effects describe the evolution of the *normal* eukaryotic cell, i.e. one with adequate mitochondria to supply the energy needed to maintain a *maximum level of order.* In fact, the energy supply is usually so generous as to allow further development into complex, *multicellular* organisms. This is a descriptive result, stating what general principle (2) and energy condition leads to multicellular organisms. Further progress would be to find why it is so, i.e. what sequence of DNA and its foldings do the trick.

By comparison, as previously found [2], energy-deficient cells such as cancer, and prokaryotes, which lack mitochondria, develop while maintaining *minimal levels* of order and information. Ironically, these energy-deficient cancers can consume large amounts of energy. (Interestingly, although *I* is minimal for such prokaryotes, their levels of PI and organization are very large.)

Now, either a maximum *or a minimum* does give the benefit of first-order stability to environmental perturbations. Nevertheless, 2$^{nd}$-order stability (as found above for normal cells) seems not to be obeyed by these energy-deficient systems. It may be, in fact, that such 2$^{nd}$-order stability is necessary for the further organization of cells into *permanently multicellular* organisms: neither cancers nor prokaryotes organize into ordered, permanently multi-celled organisms (some prokaryotes are *temporarily* multi-celled, as a stage of development). (Note: Of course the word 'order' in '2$^{nd}$-order' does not refer to the order *R* of a system.) Also, by order *R* we mean measure (18), and not "organization" in the sense of PI [27].

A point of possible confusion is that principle (2) of *I = max*. leads to *self-ordered* systems, rather than "self-organized" ones. By comparison, "self-organization" is without empirical and prediction-fulfilling support. No falsifiable theory of self-organization exists. "Self-organization" provides no mechanism and offers no detailed verifiable explanatory power. Instead it requires decisions to be provided by outside mechanisms, e.g. by configural switching.

## 11. ACKNOWLEDGMENTS

The authors acknowledge support from the National Cancer Institute under grant 1U54CA143970-01. We thank M. Subramanian for clarifying the roles played by double potential wells in fostering system emergent properties.

**APPENDIX A: Derivations and meanings of Fisher information**

Fisher information arises, independently, as both as a measure of system order and a measure of the quality of its data. It also defines a distinct arrow of time (possibly the 'biological' one) and may be used as well to derive all laws of *inanimate* physics. Hence there are two independent derivations of Fisher information, depending upon the problem – whether system order or data quality -- that is addressed.

1. *Fisher information as a measure of order*

The first derives Fisher information *I* as a measure of order (or complexity). Let the system be defined by its probability amplitude law *q*(**x**), **x** = $x_{1,...}x_K$. The corresponding probabilities are *p*(**x**) = $q^2$(**x**). Order can only be a property of a finite system. This is in view of the 2$^{nd}$ law of thermodynamics, which requires disorder to increase over *all* space. Hence a system of finite extent can have increased order, providing a compensating level of disorder is shipped outside it. Thus, increased order can only exist in finite 'pockets' of space.

Hence, the system *q*(**x**) must be of finite maximum extension. Let this be length *L.* Suppose that the system is randomly degraded such that it loses some (or all) of its fine structure. This is called a process of 'coarse graining,' a term that dates from when in conventional photography a fine- grained film is replaced by a coarse-grained one. The picture becomes 'grainy,' or 'snowy' in the case of television. One would then expect a coarse graining to cause a loss of system order. Hence we quantify this by the statement that under any coarse graining perturbation δ*q*(**x**) the resulting change in order δ*R*[*q*(**x**)] ≤ 0. Expanding the latter in Taylor series and using a theorem of Cencov [28] gives [16] the order as

$$R[q(\mathbf{x})] = (1/8)L^2 I, \qquad (A1)$$

where *I* is the Fisher information Eq. (1).

2. *Fisher information as a measure of the quality of data from a system*

The second Eq. (1), $e_{min} = \frac{1}{\sqrt{I}}$, shows that the larger *I* is the smaller is the minimum possible root-mean squared error $e_{min}$. This derives from the following scenario [25]. Suppose that data **y** = $y_1,...,y_N$ are known as *N* independent measurements of an unknown system parameter *a.* The parameter is to be estimated from the data. For this purpose an estimation function *a'*(**y**) of *a* is formed from the data. An example is the sample mean *a'*(**y**) = $\frac{1}{N}\sum_{n=1}^{N} a_n$. What is the smallest possible root-mean square error $e_{min}$ that any estimation function can achieve?

It is assumed that the likelihood law $p(\mathbf{y}|a)$ of the system is known (Note: The line | means "if"). Assume that the estimation function is unbiased, i.e. incorrect for any one set of data, but the average such estimate over a large number of sets of data **y** is correct, i.e. value *a.* This may be expressed as the condition

$$<a'(\mathbf{y})> = \int d\mathbf{y}\ a'(\mathbf{y}) p(\mathbf{y}|a) = a. \qquad (A2)$$

Because the likelihood law has unit area under it, this can be rewritten as

$$<a'(\mathbf{y}) - a> = \int d\mathbf{y}\ [a'(\mathbf{y}) - a]p(\mathbf{y}|a) = 0. \tag{A3}$$

The rest is mere algebra. One differentiates both sides $\partial/\partial a$, uses the identity relating the derivative of the $\ln p$ to the derivative of $p$ itself, etc., and finally the Schwarz inequality, to give the result

$$e^2 I \geq 1, \tag{A4}$$

where

$$e^2 = \int d\mathbf{y}\ [a'(\mathbf{y}) - a]^2 p(\mathbf{y}|a) \text{ and } I = \int d\mathbf{y}\ \langle \left(\frac{\partial \ln p(\mathbf{y}|a)}{\partial a}\right)^2 \rangle \tag{A5}$$

Is defined to be the Fisher information. Eq. (A4) is called the "Cramer-Rao inequality" or, simply, the "error inequality". It shows that the mean-squared error and the Fisher information obey a complementarity relation. if one is very small the other cannot also be very small. Or, taken another way, the minimum possible mean-squared error obeys

$$e_{min} = \frac{1}{\sqrt{I}}, \tag{A6}$$

which we set out to prove. Finally, assume a case of $N = 1$ data value, and where the system is shift invariant, obeying

$$p(\mathbf{y}|a) = p_X(y - a), \quad x = y - a. \tag{A7}$$

Then the 2$^{nd}$ Eq. (A5) becomes Eq. (1),

$$I = \int dx \frac{\left(\frac{dp}{dx}\right)^2}{p}, \quad p = p_X. \tag{A8}$$

The Cramer-Rao inequality (A4) has many applications to scientific laws in and of itself [2].

3. *Fisher information and order R define arrows of time*

Suppose that a system is perturbed by being coarse-grained (Sec. 6.4). This 'lossy' process causes tiny losses [16] $\delta I$ and $\delta R$ in Fisher $I$ and order $R$,

$$\delta I \leq 0, \text{ and } \delta R \leq 0 \text{ for } dt > 0. \tag{A9}$$

That is, during any tiny passage $dt$ of time both the information $I$ and the order $R$ must either suffer a decrease or stay the same if the system is coarse grained (i.e., loses structural detail). Thus, if the system is forced by some outside influence to lose structural detail it will tend to lose both order and information.

4. *Fisher information is used in a principle that generates the physical laws of inanimate systems*

The principle of Extreme physical information (EPI) has been used to derive most textbook physics and certain laws of biology and economics [6a]. This principle has two requirements: $I - J$ = extreme value, and $I = \kappa J$, where $\kappa \leq 1$. In this principle, $I$ always has the functional form Eq. (1). That is, $I$ is regarded as generic information. By comparison, the new information quantity $J$ is the physical expression of $I$ within the system (before it is observed). This requires some physical knowledge of the system (say, an equation of continuity). Since any observation incurs loss of information, the maximum possible value of $I$ is $J$, and this is why information efficiency $\kappa \leq 1$ in the above. Solution to the first requirement $I - J$ = extreme value is by use of the calculus of variations for a continuous system, or by ordinary differential calculus for a discrete system.

## APPENDIX B: Derivation of flux *F* curve in Figs. 1,2

A messenger protein of mass *m* on its trajectory from CM to NPC is subjected to two forces [8a]:

(1) A Coulomb force of attraction $F_C(r) = -zqE(r)$ *toward* the NPC, where electrostatic field
$$E(r) = \left(\frac{Q_{NM}}{4\pi\varepsilon r^2}\right)\left[\left(\frac{1+k_0 r}{1+k_0 a}\right)e^{-k_0(r-a)}\right]. \tag{B1}$$
A specific charge of *z=2* electron charges *q* is assumed. The Debye-Huckel constant $k_0$ obeys Eq. (4) of the text.

(2) A drag force, *away* from the NPC,
$$F_D(r) = -K dr/dt. \tag{B2}$$
(All constants in these equations are defined in Table 1.)

By Newton's 2nd law the total force $F = m(d^2r/dt^2) = F_C(r) + F_D(r)$. Using Eqs. (B1) and (B2), and noting that the acceleration force $m(d^2r/dt^2)$ is negligible compared to the other two, gives a net result
$$zqE(r) + K dr/dt = 0. \tag{B3}$$
This is in the familiar 'terminal velocity' *dr/dt* form, although here it varies in *r* according to the field *E(r)*. Substituting in Eq. (B1) gives a first-order differential equation
$$dt = -B \frac{r^2}{1+k_0 r} e^{k_0 r} dr, \text{ with } B = -K\left(\frac{4\pi\varepsilon}{zqQ_{NM}}\right)\left(\frac{1+k_0 a}{\exp(k_0 a)}\right) = const. \tag{B4}$$
Both sides of the 1st equation may be analytically integrated, the left side from *t = 0* to a general time *t,* the right side from initial position $r_0$ to general position *r*. However, the latter integral is not in a form directly found in the tables. But the change of variable $1 + k_0 r = -x$ puts it in the form
$$\int_{-1-k_0 r_0}^{-1-k_0 r} dx \left(x + 2 + \frac{1}{x}\right) e^{-x}. \tag{B5}$$
This is the sum of 3 integrals, of which the first 2 are elementary. The last brings in the exponential integral function, as $E_1(1 + k_0 r) - E_1(1 + k_0 r_0)$. (Note: $E_1$ is denoted as *Ei* in some texts, e.g. as computed by Scientific Workplace to produce Figs. 1 and 2.)

Integrating the left side of (A4) gives simply *t* as, now, a known function of *r* (right side). Evaluating the latter at *r=a* defines the function $t(a) = t_a$ by definition. This function is then substituted into the relation $F = \rho(r_0 - a)/t_a$ found above Sec. 4.1. Since both ρ (by Eq. (4)) and $t_a$ are now known functions of $k_0$ this allows the function $F(k_0)$ to be computed. The time $t_a$ is found to be about 0.016s, surprisingly fast. However this, in fact, is consistent with clinical data, where cell response times of from 0.01s – 0.1s were required following trauma injury [8b].

## APPENDIX C : Transition from K-L entropy to Fisher measure within nucleus

The aim is to prove the transition (23) from K-L entropy to Fisher information. Whereas Eq. (1) defines the Fisher *I* for a *continuous* variable *x,* our case is that of a discrete variable *i*. The definition for this case is [1,2]
$$I = \sum_{bi} \frac{(P_{b,i+1} - P_{bi})^2}{P_{bi}} \tag{C1}$$

where $P_{bi} = P_b(x_i) = P_b(i\Delta x)$ with *i* the location number along the DNA lattice. Factoring $P_{bi}$ in the numerator of (C1) gives

$$I = \sum_{bi} P_{bi} \nu^2, \quad \nu = \left(\frac{P_{b,i+1}}{P_{bi}} - 1\right). \tag{C2}$$

Now assume that $P_{bi}$ is a smooth function of *i*, so that $P_{b,i+1} = P_{bi} + \Delta P_{bi}$ with $\Delta P_{bi} \to 0$. Then ν in (C2) is close to zero. But by Taylor series $\ln(1 + \nu) = \nu - \nu^2/2$, or equivalently

$$\nu^2 = 2[\nu - \ln(1 + \nu)] \tag{C3}$$

Then (C2) becomes

$$I = 2\sum_{bi} P_{bi} \left(\frac{P_{bi} + \Delta P_{bi}}{P_{bi}} - 1\right) - 2\sum_{bi} P_{bi} \ln\left(\frac{P_{bi} + \Delta P_{bi}}{P_{bi}}\right). \quad (C4)$$

The first right-hand term is identically zero since $\sum_{bi} \Delta P_{bi} = \Delta \sum_{bi} P_{bi} = 0$ by normalization of probabilities $P_{bi}$. This leaves

$$I = 2 \lim_{\Delta P_{bi} \to 0} \sum_{bi} P_{bi} \ln\left(\frac{P_{bi}}{P_{bi} + \Delta P_{bi}}\right) = 2H(P,Q) \quad (C5)$$

by Eq. (23). QED

**APPENDIX D: Deriving linear relation between information *I* and flux *F***

In general, the flux of particles is the number of particles per unit area per unit time. The average flux of proteins from CM to NPC obeys

$$F \equiv \rho <v> = N/(t_a A). \quad (D1)$$

This describes N proteins traversing the CM-NPC over time $t_a$, where $A$ is the cross sectional area of the NM. We assume that the protein positions $y$ are processed by the NM so as to estimate the ideal position NPC $x_0$. The maximum likelihood estimate is then just the arithmetic mean of the total excursions $y$. The mean-squared error in a single excursion is $<x^2>$, and its information is $I_1 = 1/<x^2>$ after use of definition (1) where $p(x)$ is assumed to be Gaussian with variance $<x^2>$. Then since the protein fluctuations $x$ are independent, their informations add [25] and the total $I = N I_1 = N/<x^2>$. The molecules of the cytoplasm induce random walk in the proteins, so that their mean-squared position at the NPC $<x^2> = 2Dt_a$ with $D$ the diffusion constant. Thus the information in $N$ readings deliver a total information $I = N/<x^2> = N/(2Dt_a)$. Using this in Eq. (D1) gives

$$I = \left(\frac{A}{2D}\right) F. \quad (D2)$$

**CORRESPONDING AUTHOR INFORMATION**

Tel: (520) 621-4904
Fax: (520) 621-3389
E-mail: roy.frieden@optics.arizona.edu

**RUNNING TITLE**

Cell development obeys maximum Fisher information